\documentclass[prb,aps,twocolumn,superscriptaddress,10pt]{revtex4-1}
\usepackage{amsmath}
\usepackage{amssymb}
\usepackage{amsthm}
\usepackage{amsfonts}
\usepackage{listings}
\usepackage{enumerate}
\usepackage{latexsym}
\usepackage{psfrag}
\usepackage{bm}
\usepackage[all]{xy}
\usepackage{graphicx}
\usepackage{subfigure}
\usepackage[pdftex,colorlinks=false]{hyperref}
\usepackage{color}
\usepackage{mathtools}
\usepackage{eufrak}

\begin{document}

\title{Gapped spin liquid with $\mathbb{Z}_2$-topological order for Kagome Heisenberg model}

\author{Jia-Wei Mei}
\affiliation{Perimeter Institute for Theoretical Physics, Waterloo, Ontario, N2L 2Y5, Canada}
\author{Ji-Yao Chen}
\affiliation{State Key Laboratory of Low Dimensional Quantum Physics, Department of Physics, Tsinghua University, Beijing 100084, China}
\author{Huan He}
\affiliation{Department of Physics, Princeton University, Princeton, New Jersey 08544, USA}
\author{Xiao-Gang Wen}
\affiliation{Department of Physics, Massachusetts Institute of Technology, Cambridge, Massachusetts 02139, USA}
\affiliation{Perimeter Institute for Theoretical Physics, Waterloo, Ontario, N2L 2Y5, Canada}

\date{\today}

\begin{abstract}
We apply symmetric tensor network state (TNS) to study the nearest neighbor spin-1/2 antiferromagnetic Heisenberg model on Kagome lattice. Our method keeps track of the global and gauge symmetries in TNS update procedure and in tensor renormalization group (TRG) calculation. We also introduce a very sensitive probe for the gap of the ground state -- the modular matrices, which can also determine the topological order if the ground state is gapped. We find that the ground state of Heisenberg model on Kagome lattice is a gapped spin liquid with the $\mathbb{Z}_2$-topological order (or toric code type), which has a long correlation length $\xi\sim 10$ unit cell length. We justify that the TRG method can handle very large systems with over thousands of spins. Such a long $\xi$ explains the gapless behaviors observed in simulations on smaller systems with less than 300 spins or shorter than 10 unit cell length. We also discuss experimental implications of the topological excitations encoded in our symmetric tensors.
\end{abstract}
\maketitle

%\tableofcontents

\section{Introduction}\label{sec.Introduction}

The pattern of long-range entanglement defines topological orders in gapped quantum phases of matter that lie beyond the Landau symmetry breaking paradigm\cite{Wen2004}. Quantum spin liquid (QSL) state is a concrete example for topologically ordered states\cite{Anderson1987,Wen1991,Wen2002,Lee2008,Balents2010,Sheng2014,Sheng2015,HeShengYan2014,Bauer2014}.
In (2+1) dimension, topological order is described by new topological quantum numbers, such as non-trivial ground state degeneracy and fractional excitations\cite{Laughlin1983,Wilczek1984,Arovas1984,Wen1989,Wen1990,Wen1990a}. These topological properties are fully characterized by modular matrices for the degenerate ground states\cite{Keski-Vakkuri1993,Wen1990a,Wen1989,Wen1990,Wen2012,Liu2013,Moradi2014,He2014,Mei2015},
whose elements encode the mutual statistics and topological spins of excitations\cite{Wen01032016}. 

Generally, a many-body quantum state is expressed as
\begin{equation}\label{eq:ltensor}
|\psi\rangle=\sum_{s_1,\cdots,s_N}T^{s_1,...,s_N}|s_1,\cdots,s_N\rangle, 
\end{equation}
where the coefficient $T^{s_1,...,s_N}$ can be viewed as a tensor exponentially large with system size. 
For a gapped local Hamiltonian, 
entanglement entropy of its ground states typically obeys area law\cite{Eisert2010}, thus $T^{s_1,...,s_N}$ can be approximated by contractions of small local tensors. A set of variational states have been proposed, such as matrix product states in one dimensional and quasi two dimensional (2D) systems\cite{Schollwock2010}, projected entangled pair states in 2D systems\cite{Verstraete2004} and other types of tensor network state (TNS)\cite{Shi06,Vidal08}. 
For TNS that describes a topologically ordered wave function in 2D, the gauge symmetry of local tensors is necessary\cite{ChenZengGuChuangWen}, i.e., each local tensor 
should be invariant under local symmetry transformations on virtual legs. For a gauge symmetric TNS, we are able to compute modular matrices\cite{Moradi2014,He2014,Mei2015} which can detect topological phase transitions\cite{Neupert2016,Neupert2016Nogo,Lan2015}.

The nearest neighbor (NN) spin-1/2 Kagome antiferromagnetic Heisenberg model (KAFHM),
\begin{equation}\label{eq:KHM}
H=\sum_{\langle ij\rangle}\mathbf{S}_i\cdot\mathbf{S}_j,
\end{equation} 
is thought to host a QSL ground state. ZnCu$_3$(OH)$_6$Cl$_2$ [herbertsmithite]\cite{Shores2005,Helton2007,Mendels2007,Zorko2008,Imai2008,Vries2009,Imai2011,Jeong2011,Han2012a,Han2012,Han2016,Fu2015} and Cu$_3$Zn(OH)$_6$FBr\cite{Feng2017} are promising compounds for experimental realizations of KAFHM with additional interactions.
Many different ground states have been proposed for KAFHM\cite{Marston1991,Sachdev1992,Mila1998,Hastings2000,Nikolic2003,Budnik2004,Wang2006,
Ran2007,Singh2008,Hermele2008,Jiang2008,Evenbly2010,Poilblanc2010,Lu2011,Schuch2012,
Poilblanc2013,Iqbal2013,Xie2014,Jiang2015,Gotze2011,Yan2011,Jiang2012,Depenbrock2012,Jiang2016,Liao2016,He2016,changlani2017mother}. 
 Numerically, the initial density matrix renormalization group
(DMRG) calculations\cite{Yan2011,Jiang2012,Depenbrock2012} supported a
symmetric $\mathbb{Z}_2$ QSL ground state. Considering time reversal
symmetry and translational symmetries, Zaletel \textit{et al}\cite{ZaletelVishwanath2015} argued that the $\mathbb{Z}_2$ QSL should be
$\mathbb{Z}_2$-gauge type\cite{RS9173,W9164} ({\it i.e.} toric code type\cite{K032}).
However, DMRG simulations fail to find all four degenerate ground states on a torus\footnote{Private communications with Lukasz Cincio and Steve White.}, or braiding statistics of quasiparticles for the
$\mathbb{Z}_2$-topological order\cite{RS9173,W9164}. 

In this paper, we use TNS to address the (topological) nature of the KAFHM ground state. 
Since previous studies did not detect any symmetry breaking order, we will assume that the KAFHM ground state can be described by a TNS with translational, time reversal and $\mathrm{SU}(2)$ spin-rotational symmetries. SU(2) symmetry implies a $\mathbb{Z}_2$ gauge symmetry on the TNS\cite{Jiang2015} and then we can study both $\mathbb{Z}_2$ topologically ordered and trivial states\cite{He2014}. We stress that without implementation of the $\mathbb{Z}_2$ gauge symmetry, it would be hard to identify the topological order in the TNS. 

We introduce the modular matrices as a very sensitive probe for the gap of the ground state and use it to determine the topological order for a gapped ground state. The tensor renormalization group (TRG) flow of modular matrices have very different behaviors for gapped and gapless ground states\cite{He2014}. We find that the KAFHM ground state is a gapped QSL with the $\mathbb{Z}_2$-topological order\cite{RS9173,W9164} and has a long correlation length, $\xi\sim10$ unit cells. We justify the TRG flow of modular matrices at critical points and argue that our estimation of the correlation length is valid within a distance of 30 unit cell length or more, consistent with previous studies\cite{GW0931,YW151204938}. Such a long correlation length ($\xi\sim$ 10 Kagome unit cells) might explain the DMRG's failure in identifying $\mathbb{Z}_2$ topological order and the gapless behaviors in recent numerical simulations\cite{Jiang2016,Liao2016,He2016,lauchli2016s,changlani2017mother}. We also discuss low energy excitations in different topological sectors and their experimental implications.

The organization of the rest of the paper is as follows: 
In Sec.~\ref{sec.TNS}, we introduce a set of $\mathrm{SU}(2)$ symmetric TNS states.
In Sec.~\ref{sec.update}, we present our algorithm to find the variational ground states given by the symmetric TNS.
In Sec.~\ref{sec.TRG}, we show how to compute TRG without breaking $\mathrm{SU}(2)$ symmetry, which is useful to compute modular matrices.
In Sec.~\ref{sec.modmat}, we present the results of the modular matrices which are unique and complete signatures of topological phases. Moreover, from the convergence speed of modular matrices with respect to TRG steps, we can estimate the correlation length of the TNS wave function.  More importantly, the method is justified by comparing with the results of critical systems. Hence, it is safe to state that this method is sensitive to the gapped/gapless nature of the ground state wave function.
In Sec.~\ref{sec.energy}, we explain our method to calculate the ground state energy, and present the energy with different bond dimension.
In Sec.~\ref{eq.discussion}, we conclude the paper by summarizing and discussing the possible implications for experiments.

A few more appendices are attached for better explanations:
In App.~\ref{app.symmetries}, we explain the techniques of keeping symmetries with more details.
In App.~\ref{app.update}, we briefly review the simple update algorithm.
%In App.~\ref{app.correlation}, we show the modular $S$ matrices for each step of TRG.

\begin{figure}[b]
  \centering
  \includegraphics[width=\columnwidth]{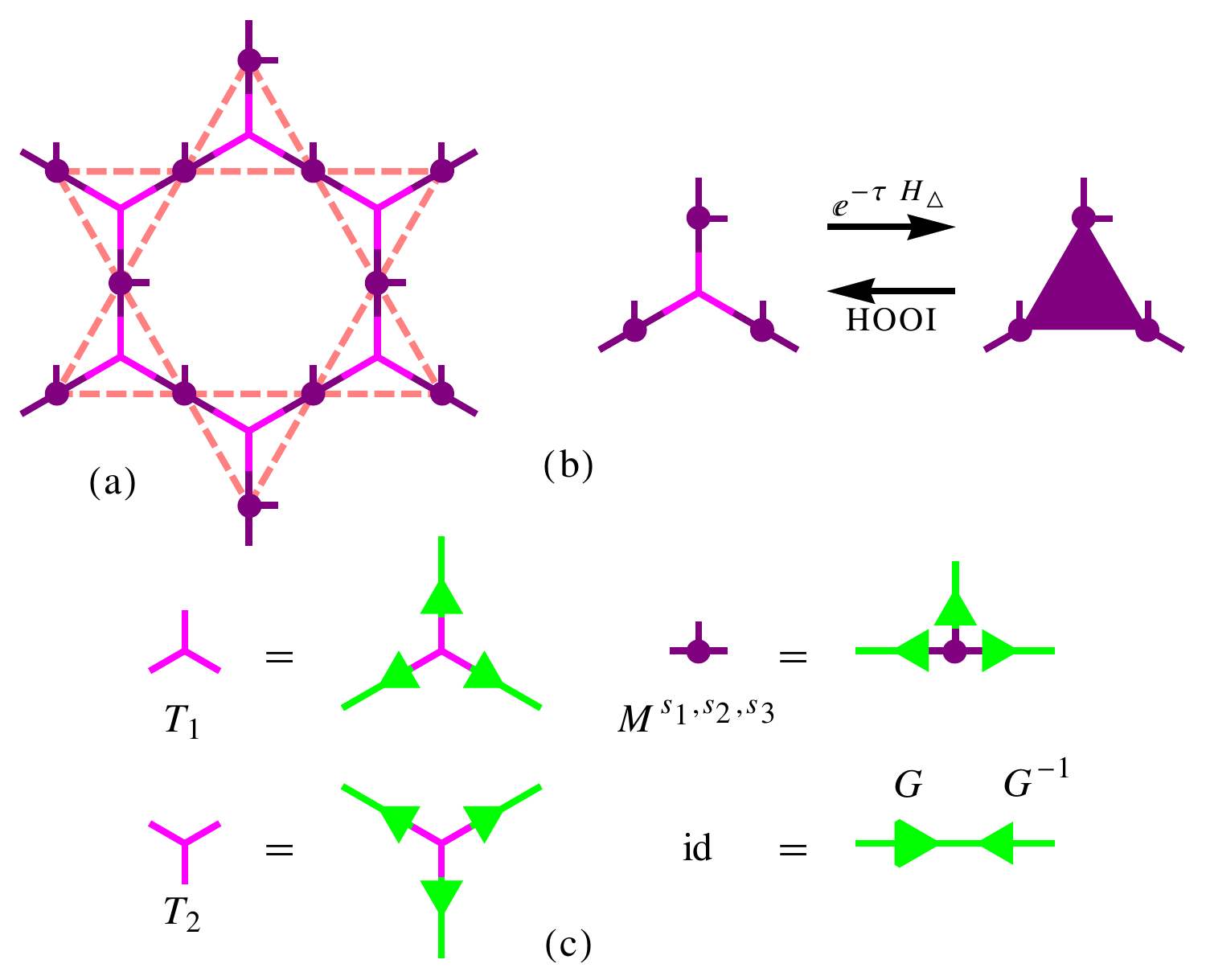}
  \caption{
    (a) TNS on Kagome lattice. Darker purple tensors are site tensors $M^{s_1,s_2,s_3}$ carrying physical spins, while lighter purple $T_{1,2}$ are joint tensors.
    (b) Simple update for TNS in which we use HOOI\cite{Lathauwer2000,Kolda2009} to truncate $\exp(-\tau H_{\triangle}) T_1M^{s_1}M^{s_2}M^{s_2}$ back to $T_1$ and $M^{s_1,s_2,s_3}$. 
    Similarly for downward triangles. (c) Symmetry condition for the local tensors. Green tensors are the symmetry matrices $G^i$'s in Eq.~\eqref{eq:symm}. `id' stands for identity.
  }
  \label{fig:TNS}
\end{figure}

\section{Symmetric TNS}\label{sec.TNS}

We express the exponentially large tensor
$T^{s_1,...,s_N}$ on Kagome lattice as contractions of the local
tensors\cite{Xie2014}
\begin{eqnarray}\label{eq:TNS}
T^{s_1,...,s_N}=\mathrm{tTr}\left(T_1T_2M^{s_1}M^{s_2}M^{s_3}\cdots\right),
\end{eqnarray}
as shown in Fig.~\ref{fig:TNS} where $M^{s_1,s_2,s_3}$ are the site tensors that contain the physical spin-1/2 legs forming the Kagome lattice and $T_{1,2}$ are the three-legs joint tensors; $\mathrm{tTr}$ denotes the tensor contraction over connected legs. We require $\mathrm{SU}(2)$ spin rotation, time reversal and translation symmetries to search the possible topological ground states. 

$\mathrm{SU}(2)$ spin rotation symmetry requires that local tensors are invariant under symmetry operators:
\begin{equation}\label{eq:symm}
\begin{split}
M^{s_i} &= \mathrm{tTr}\left((G^{s_i}(\vec{\theta})\otimes G^{i}(\vec{\theta})\otimes G^{j}(\vec{\theta})) M^{s_i}\right),	\\
T_{1,2} &= \mathrm{tTr}\left(T_{1,2} (G^{i}(-\vec{\theta})\otimes G^{j}(-\vec{\theta})\otimes G^{k}(-\vec{\theta}))\right),
\end{split}
\end{equation}
where $\vec{\theta}$ is the angle of global $\mathrm{SU}(2)$ rotation and $G^{i}$ are matrices of (projective) representations. Since $\mathrm{SU}(2)$ has no projective representations, every virtual leg must obey (reducible) representations of $\mathrm{SU}(2)$, i.e.: %local tensors obey the tensor product of these representations:
\begin{equation}\label{eq:V}
\begin{split}
M^{s_i}&\in\bigoplus_{i,j\in J}\mathcal{V}_{\frac{1}{2}}\otimes\mathcal{V}_i\otimes\mathcal{V}_j,	\\
T_{1,2}&\in\bigoplus_{i,j,k\in J}\mathcal{V}_i\otimes\mathcal{V}_j\otimes\mathcal{V}_k,
\end{split}
\end{equation}
where $\mathcal{V}_{i}$'s are irreducible representations of $\mathrm{SU}(2)$ with spin-$i$ and $J$ is a collection of spins. The physical leg is associated with a 2-dimension Hilbert space while the virtual legs have the multiplet dimension:
\begin{equation}
D^*=\sum_{j\in J}1,
\end{equation}
dubbed in Refs.~\onlinecite{SymmetryWeichselbaum,symmetryLi,SymmetryLiu}. The corresponding bond dimension is:
\begin{equation}
D=\sum_{j\in J}(2j+1).
\end{equation}

Subtly, the $2\pi$ $\mathrm{SU}(2)$ global spin-rotation transformation implies $\mathbb{Z}_2$ gauge symmetry in TNS, i.e., local tensors are invariant (up to some constant) under $\mathbb{Z}_2$ gauge transformations. For adiabatic global $\mathrm{SU}(2)$ spin rotation $0\rightarrow2\pi$ for certain ground state, the system goes back to the ground state. The matrix representation of $2\pi$ global $\mathrm{SU}(2)$ spin rotation acting on a virtual leg is $\bigoplus_{j\in J} (-)^{2j}\mathbb{I}_{2j+1}$, which is $\mathbb{Z}_2$ gauge symmetry in the center of $\mathrm{SU}(2)$ representation. $\mathbb{Z}_2$ gauge symmetry in 
TNS does not always indicate $\mathbb{Z}_2$ topological order. Only when $\mathbb{Z}_2$ gauge symmetry deconfines, $\mathbb{Z}_2$ topological order survives, i.e. the system goes to another ground state after adiabatic global spin rotation $0\rightarrow2\pi$. 

All the tensors are real and thus satisfy time reversal symmetry. The lattice symmetries (e.g. translation symmetry) can be extended by $\mathbb{Z}_2$ invariant gauge group (IGG) and the projective representations are classified by the projective symmetry group (PSG) method\cite{Wen2002,Wang2006,Jiang2015}. In this paper, we focus on one particular PSG class in which the translation symmetry has a trivial projective representation and all tensors are generated by translation transformations on tensors $T_{1,2}$ and $M^{s_1,s_2,s_3}$, which is further discussed in Sec.~\ref{sec.update}.

We stress that keeping track of gauge symmetry throughout the entire computation is essential to generate certain topological states in 2D TNS. Once the IGG of 2D TNS is known, a set of basis for the degenerate ground states can be obtained by gauge transformations\cite{SwingleWen,SchuchCirac,Buerschaper}. One may naively expect that the gauge symmetry on the virtual legs should automatically emerge after the numerical optimizations. However, this is not always true for 2D topologically ordered TNS due to the following two reasons: (1) The gauge symmetry in 2D TNS is a necessary condition for topological order at thermodynamic limit\cite{ChenZengGuChuangWen}. Adding small perturbation to the tensor that violates the gauge symmetry will completely destroy the topological order\cite{ChenZengGuChuangWen}. (2) Balents\cite{Balents} pointed out that on an infinite lattice the variational energy density of a topological ordered state can be approximated arbitrarily close by a state in a different topological class. Therefore, gauge symmetry is crucial to determine the phase of KAFHM.

\section{Symmetric Simple Update}\label{sec.update}

We develop an intuitive and simple method to keep track of symmetries during numerical truncated singular value decomposition (SVD), whose details can be found in App.~\ref{app.symmetries}. It can be easily integrated in several TNS algorithms, including imaginary time evolution and TRG. In practice, we find it also efficient. In this section, we explain how to use it to keep symmetries in simple update algorithm to find the ground states.
 
We apply the imaginary-time evolution operator $\exp\left(- \tau H \right)$ on an \emph{initial symmetric} TNS to approach the ground state in the limit $\tau \rightarrow \infty$.
We increase the multiplet dimension $D^*$ from $D^*=2$ ($D=3$) gradually in symmetric update. For $D^*=2$, the vector space of virtual legs is composed of spin-0 and spin-1/2, $\mathcal{V}_{D^*=2}=\mathcal{V}_{j=0}\oplus\mathcal{V}_{j=1/2}$ and TNS can describe four different PSG classes of nearest neighbor (NN) resonating-valence-bond (RVB) states with $\mathbb{Z}_2$ toric code topological order\cite{Wang2006,Yang2012,Jiang2015}. The PSG classification is also valid for large bond dimensions\cite{Jiang2015} where further RVB bonds longer than NN ones are involved. In this paper, we will focus on one particular PSG class, named as $Q_1=Q_2$ state in Ref.~\onlinecite{Sachdev1992} which has been suggested as a promising ground state\cite{Lu2011}. The translation symmetry for $Q_1=Q_2$ state has trivial PSG in TNS.

In the update procedure, we use our method in App.~\ref{app.symmetries} to preserve symmetries. The bond dimension of TNS should be selected carefully and we use $2\pi$ $\mathrm{SU}(2)$ rotation symmetry to remove round-off errors. $2\pi$ $\mathrm{SU}(2)$ rotation symmetry is diagonal matrix with diagonal elements $1$ for integer $j$ and $-1$ for half-integer $j$. In every step of symmetric update, $2\pi$ $\mathrm{SU}(2)$ rotation matrix can be obtained. We set the elements to exact integers $0$ (off-diagonal), $-1$ and $1$ (diagonal) to remove round-off error and then symmetrize the tensors. 

The above symmetric update can be easily implemented in simple update\cite{Jiang2008a,Xie2014}, cluster update\cite{Wang2011} and full update\cite{Jordan2008}. In this paper, we utilize the simple update algorithm\cite{Xie2014} to approach the ground state illustrated in Fig.~\ref{fig:TNS}. Besides the symmetrization, we use high-order-orthogonal-iteration (HOOI) method\cite{Kolda2009,Lathauwer2000} instead of high order SVD (HOSVD) as in Ref.~\onlinecite{Xie2014}. If we keep multiplet dimension $D^*=2$ fixed, we find that $Q_1=Q_2$ NN RVB state is the fixed point TNS after many successive imaginary-time evolution steps. This is very different from non-symmetric simple update\cite{Xie2014} where NN RVB state is not the fixed point TNS. More simple update details can be found in the App.~\ref{app.update}.

Here we list the spin collection on virtual legs for different multiplet dimension $D^*$
\begin{equation}\label{eq:Japp}
\begin{split}
J_{D^*=2}	=&	\{0,\frac{1}{2}\},	\\
J_{D^*=3}	=&	\{0,\frac{1}{2},1\}, 	\\
J_{D^*=4}	=&	\{0,\frac{1}{2},1,\frac{1}{2}\},	\\
J_{D^*=5}	=&	\{0,\frac{1}{2},1,\frac{1}{2},0\},\\
J_{D^*=6}	=&	\{0,\frac{1}{2},1,\frac{1}{2},0,\frac{3}{2}\}	\\
J_{D^*=7}	=&	\{0,\frac{1}{2},1,\frac{1}{2},0,\frac{3}{2},1\} \\
J_{D^*=8}	=&	\{0,\frac{1}{2},1,\frac{1}{2},0,\frac{3}{2},1,\frac{1}{2}\},	\\
J_{D^*=12}	=&	\{0,\frac{1}{2},1,\frac{1}{2},0,\frac{3}{2},1,\frac{1}{2},1,\frac{1}{2},0,2\},	\\
J_{D^*=20}	=&	\{0,\frac{1}{2},1,\frac{1}{2},0,1,\frac{3}{2},\frac{1}{2},\frac{1}{2},1,	\\
&~~ 0,0,\frac{3}{2},1,\frac{1}{2},2,\frac{1}{2},1,\frac{3}{2},\frac{1}{2}\}.
\end{split}
\end{equation}

\section{Symmetric TRG  and Modular matrice evaluations}\label{sec.TRG}
\begin{figure}[t]
  \centering
  \includegraphics[width=\columnwidth]{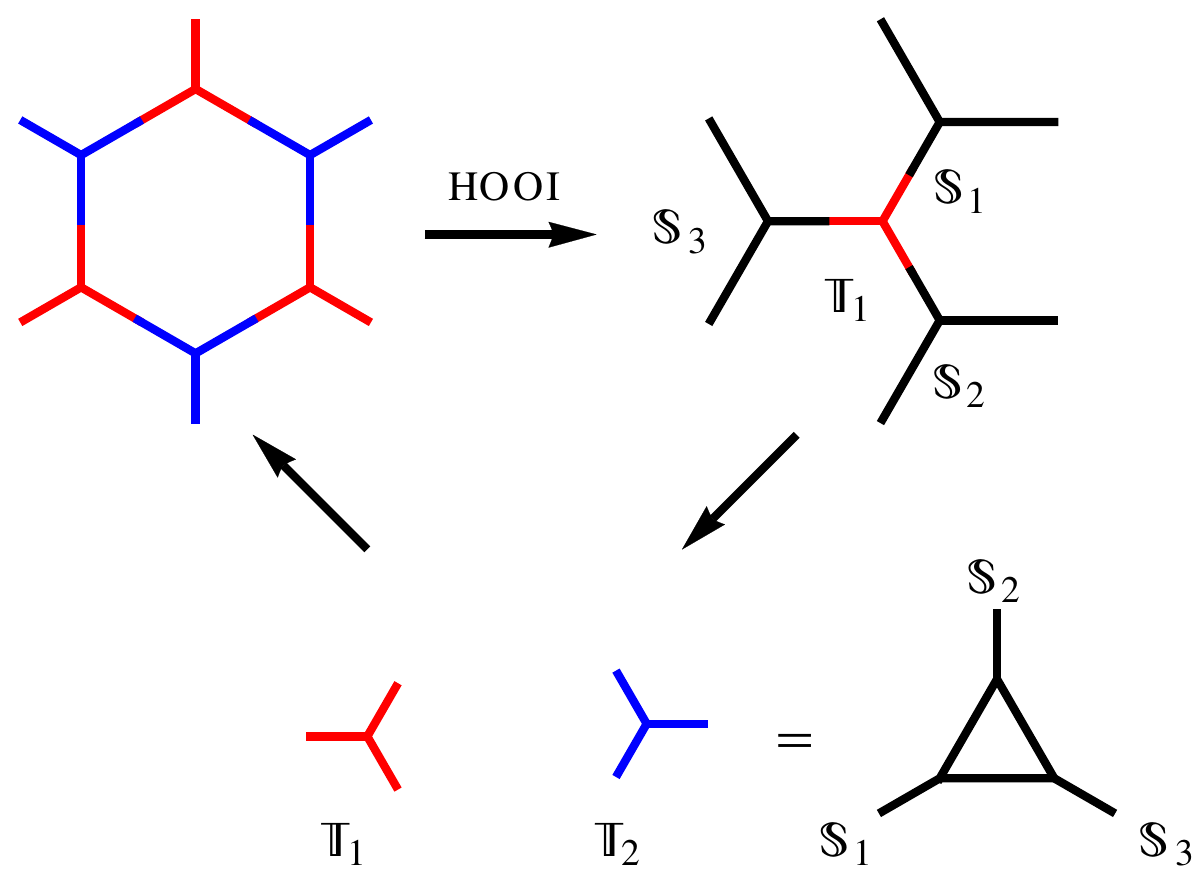}
  \caption{
  TRG diagrammatic scheme. The original honeycomb lattice can be taken as a \emph{triangular} lattice in which the unit cell is the six-leg tensor $\mathcal{T} \equiv \mathrm{tTr} \left( \mathbb{T}_1 \mathbb{T}_2 \mathbb{T}_1 \mathbb{T}_2 \mathbb{T}_1 \mathbb{T}_2 \right)$ as shown on the left.  The triangular lattice TN can be deformed into a TN on honeycomb lattice\cite{Gu2008}. 
  }
  \label{fig:TRG}
\end{figure}
Given a TNS, we need to evaluate the norm of the wave function:
\begin{eqnarray}\label{eq:norm}
\langle\psi|\psi\rangle=\mathrm{tTr}[\mathbb{T}_1\mathbb{T}_2\cdots],
\end{eqnarray}
where $\mathbb{T}_{1,2}$ are coined ``double tensors" defined as follows:
\begin{equation}\label{eq:doubleT}
\begin{split}
\mathbb{T}_1&=\sum_{s1,s2,s3}(\text{TM}_1^{s_1s_2s_3})^*\text{TM}_1^{s_1s_2s_3},	\\
\mathbb{T}_2&=T_2^*T_2,
\end{split}
\end{equation}
with $\text{TM}_1^{s_1s_2s_3}=\text{tTr}(T_1M^{s_1}M^{s_2}M^{s_3})$.

Let's first explain our TRG scheme without symmetries which is improved from TRG in Ref.~\onlinecite{Levin2007}. The TRG is shown diagrammatically in Fig.~\ref{fig:TRG}. We take the six-leg tensors $\mathcal{T} \equiv \mathrm{tTr} \left( \mathbb{T}_1 \mathbb{T}_2 \mathbb{T}_1 \mathbb{T}_2 \mathbb{T}_1 \mathbb{T}_2 \right)$ as shown on the left in Fig.~\ref{fig:TRG} as the unit cell of the next coarse-graining lattice. The honeycomb lattice turns out to be the triangular lattice composed of $\mathcal{T}$. We use HOOI to decompose the six-leg tensor $\mathcal{T}$ on the left into one core tensor (red) and three joint tensors $\mathbb{S}_{1,2,3}$ (black). The core tensor and the contracted tensor from $\mathbb{S}_{1,2,3}$ are new double tensors $\mathbb{T}_1$ and $\mathbb{T}_2$ in the next TRG step. This is similar to Fig.12 and Fig.13 in Ref.~\onlinecite{Gu2008}. In practice, we only store $\mathbb{T}_{1,2}$ and $\mathbb{S}_{1,2,3}$ in memory. During HOOI, we do not store the whole six-leg tensor $\mathcal{T}$ explicitly, but access the tensor by evaluating tensor contractions (matrix-vector multiplication) during truncated SVD. We use lmsvd\cite{Liu2013a} to do the truncated SVD in HOOI. The whole memory cost is $\mathcal{O}(\chi^3)$ and the computational cost is $\mathcal{O}(\chi^5)$ where $\chi=D^2$ is the bond dimension for double tensors. 

\begin{figure}[t]
  \centering
  \includegraphics[width=\columnwidth]{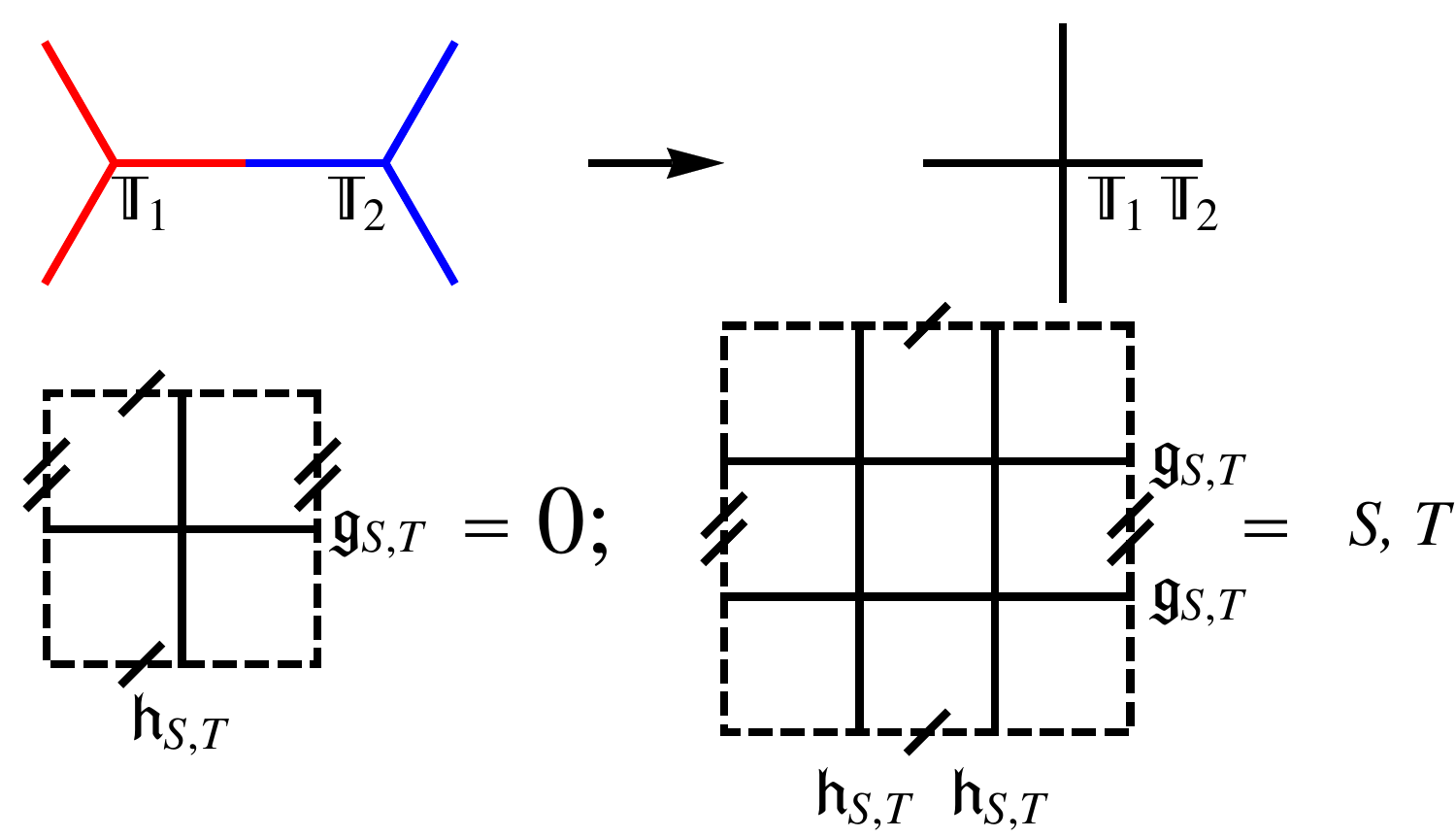}
  \caption{Modular matrices. After every step of TRG, we contract $\mathbb{T}_{1,2}$ into $\mathbb{T}_1\mathbb{T}_2$ and put it on the torus as described by dashed squares. Due to $\mathrm{SU}(2)$ symmetry, the result of contracting single $\mathbb{T}_1\mathbb{T}_2$ is zero. So we need four $\mathbb{T}_1\mathbb{T}_2$ tensors to evaluate modular matrices.}
  \label{fig:ST}
\end{figure}

Given IGG of TNS on a torus, we can insert gauge fluxes to pump the ground state from one to another, $\psi(g\rightarrow g',h\rightarrow h')$.
Now we are able to use the universal wave-function overlap method\cite{Moradi2014,He2014} to calculate the modular matrices:
\begin{equation}\label{eq:UWFO}
\begin{split}
T&=\langle\psi(g',h')|\psi(g,gh)\rangle	=\mathrm{tTr}[\mathbb{T}_1\mathbb{T}_2\circ(\frak{g}_T\frak{h}_T)\cdots],	\\
S&=\langle\psi(g',h')|\psi(h,g^{-1})\rangle	=\mathrm{tTr}[\mathbb{T}_1\mathbb{T}_2\circ(\frak{g}_S\frak{h}_S)\cdots],
\end{split}
\end{equation}
where $\frak{g}_T=g'\otimes g$, $\frak{h}_T=h'\otimes gh$, $\frak{g}_S=g'\otimes h$ and $\frak{h}_S=h'\otimes g^{-1}$. $g',h'$ and $g,h$ are in IGG acting on the bra and ket, and on the vertical and horizontal boundaries, respectively. See Fig.~\ref{fig:ST} for diagrammatic representation. After every step of TRG, we contract $\mathbb{T}_{1,2}$ into $\mathbb{T}_1\mathbb{T}_2$ and put them on the torus to evaluate the modular $S$ and $T$ matrices as illustrated in Fig.~\ref{fig:ST}. 

\section{Identify a gap from modular matrices}\label{sec.modmat}
In this section, we will show that the modular matrices is a very sensitive probe for the gap of the ground state and we can use it to determine the topological order for a gapped ground state. For gapped and gapless ground states, the TRG flow of modular matrices have very different behaviors\cite{He2014}. To justify the modular matrix method of determining gapped/gapless states, we will study the TRG flow of universal ratio $Q$ in Eq.(\ref{eq:Q}). Also from the convergence speed of the TRG flow of $Q$, we can estimate the correlation length for the gapped ground state. 

\subsection{Modular matrices for a kagome spin liquid TNS}
\label{sec:STcon}

As we have explained in Sec.~\ref{sec.TRG}, in order to compute modular matrices, we need to keep gauge symmetries during TRG. The converged modular matrices for our TNS ground states are:
\begin{equation}\label{eq:ST}
S=\left(\begin{matrix}1&0&0&0\\0&0&1&0\\0&1&0&0\\0&0&0&1\end{matrix}\right),	\quad T=\left(\begin{matrix}1&0&0&0\\0&1&0&0\\0&0&0&1\\0&0&1&0\end{matrix}\right),
\end{equation}
which are identical to modular matrices for $\mathbb{Z}_2$ toric code phase in string basis\cite{Liu2013}.
Modular matrices in Eq. (\ref{eq:ST}) provide complete information of the topological order\cite{W9039,W150605768} where the mutual statistics and topological spins of anyons can be read off from $S$ and $T$ elements, respectively. This is beyond the state-of-art DMRG computations which support $\mathbb{Z}_2$ gapped spin liquid\cite{Yan2011,Jiang2012,Depenbrock2012}, only demonstrated by topological entanglement entropy.

\subsection{TRG flow of modular matrices for a gapless state}
\label{sec:STgapless}

\begin{figure}[t]
  \centering
  \includegraphics[width=\columnwidth]{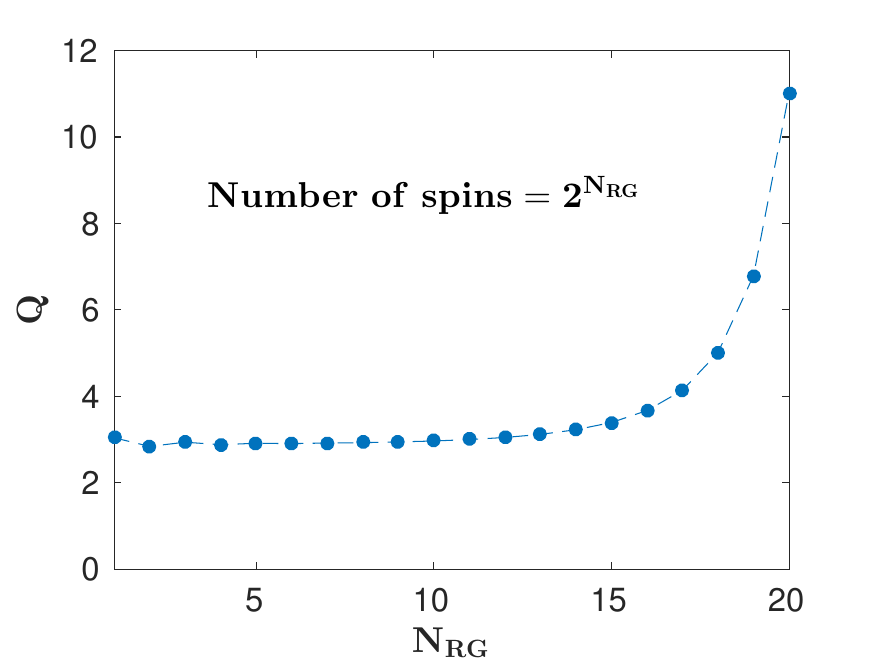}
  \caption{ The universal ratio $Q$ (Eq.(\ref{eq:Q})) v.s. number of TRG steps $N_{\text{RG}}$ for a toric code TNS at the critical point\cite{He2014}. The system has $2^{N_\text{RG}}$ spins after $N_{\text{RG}}$ steps of the TRG iteration.}
  \label{fig:QRG}
\end{figure}

%The modular matrices for our SU(2) TNS in Eq.(\ref{eq:ST}) imply a $\mathbb{Z}_2$ gapped spin liquid for the Kagome Heisenberg model. However, one may question the validity of TRG method for modular matrix evaluations. 

Since each iteration step in our calculation introduces a truncation error, if we run many steps of iterations, the truncation errors will cumulate and the result of the calculation will be dominated by the truncation errors. So we need to estimate the maximum iteration step that we can run safely before the truncation errors destroy our result. 

For this purpose, we compute the modular matrices for TNS at critical points where the truncation errors are largest. We compute the modular matrices for the ideal wave function for toric code model with string tension $g=0.802$. See the Ref.~\onlinecite{He2014} for more details about this wave function. It is actually the critical point of the condensation phase transition between toric code and trivial phase\cite{He2014}. If there was no truncation error (for the case of an infinite bond dimension), the tensor will flow to a fixed point tensor, which will give rise to the fixed point modular matrices.

To manifest the convergence or divergence of a TRG flow, we define the following universal ratio during the TRG flow:
\begin{eqnarray}\label{eq:Q}
Q=\frac{\text{trace}(S)}{|\det(S)|^{1/4}}.
\end{eqnarray}
We expect that $Q$ converges after $N_{\text{RG}}$ steps of TRG when tensors flows to fixed points.

For the critical toric code model\cite{He2014} with the truncated bond dimension $\chi=144$, the TRG flow of the universal ratio $Q$ is shown in Fig.\ref{fig:QRG}. We find that $Q$ indeed quickly approaches a constant. But after $10 \sim 12$ iteration steps, $Q$ starts to blow up. This is caused by the truncation error introduced by the finite bond dimension. So, we see for bond dimension $\chi=144$, we can perform 10 steps of iteration safely. This demonstrates the validity of TRG within ten iteration steps where the system reaches the size with $2^{10}$ spins.

\subsection{TRG flow of modular matrices for a kagome spin liquid TNS}
\label{sec:STgapless}

\begin{figure}[b]
  \centering
  \includegraphics[width=\columnwidth]{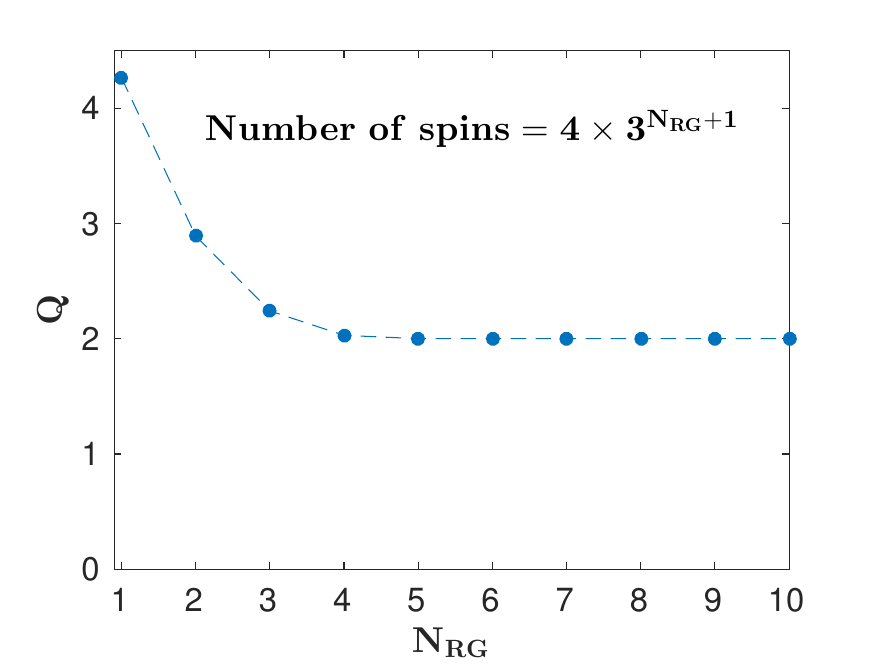}
  \caption{ The universal ratio $Q$ (Eq.(\ref{eq:Q})) v.s. number of TRG steps $N_{\text{RG}}$ for our SU(2) symmetric TNS for the kagome Heisenberg model. The system has $4\times3^{N_\text{RG}+1}$ spins after $N_{\text{RG}}$ steps of the TRG iteration.}
  \label{fig:QRGK}
\end{figure}
In the previous subsection, we justify that the TRG method can handle very large systems with over thousands of spins even for a critical state wave function. Now we are ready to implement it to study our SU(2) TNS wave function. In Fig.~\ref{fig:QRGK}, we also plot the TRG flow of the universal ratio $Q$ for our SU(2) symmetric TNS with the truncated total (not multiplet) bond dimension $\chi=140$ %\HH{Which bond dimension? Multiplet or total truncation?}. 
We see that we obtain the fixed-point modular matrices after only $3\sim 4$ steps of iterations. We still expect the TRG to be valid within ten iteration steps. So the fixed-point modular matrices obtained after 4 steps of iterations are valid reliable results. We note that at 3 steps of iterations, the modular matrices are calculated from a system of with $4\times3^{N_{\text{RG}}^c+1}=324$ spins. Therefore, we can only detect the gap in the Kagome Heisenberg model on a system with more than $300 \sim 400$ spins.
%\HH{We can safely do TRG for 10 steps. So I think the safe system size is $4*3^{10+1}=708588$. Safe TRG is determined by the converge of Q, while the unsafe TRG is determined by the divergence of Q.}

The modular matrices converge to the fixed-point value only after $3 \sim 4$ TRG steps. From the convergence speed of modular matrices, we can estimate the correlation length in our TNS, $\xi\sim 2\times3^{N^c_{\text{RG}}/2}$ in terms of the unit cell length (which is the square root of the Kagome unit cell area), where $N^c_{\text{RG}}$ is the number of TRG steps after which modular matrices converge. From the $S$ matrix data for every TRG steps, we can estimate $N^c_{\text{RG}} = 3$ and the correlation length $\xi\sim 10$ Kagome unit cell length.

One might point out that the correlation length $\xi\sim 10$ Kagome unit cell length indicates a gapless critical phase since 10 unit cell length is pretty close to infinity for most numerical simulations. But 10 unit cell length is not infinity for TRG simulation used in this paper. By studying quamtun critical
state where the TRG simulation has the worst truncation errors, Refs.~\onlinecite{GW0931,YW151204938} found that TRG simulation can handle very large system of size over 30 unit cell length, since highly acurate critical exponents were obtained from calculations on such large systems. In previous subsection, we directly justify this point via the TRG flow of modular matrices on a critical point. So the obtained long correlation length, $\xi \sim 10$ Kagome unit cells, should be valid. This also explains the gapless behaviors in recent simulations on a smaller system\cite{Jiang2016,Liao2016,He2016,changlani2017mother,lauchli2016s}. Magnetic order scaling behavior in Ref.~\onlinecite{Liao2016} might be not sensitive enough to resolve such a long correlation length. 

\section{Ground State Energy}\label{sec.energy}

\begin{figure}[t]
  \centering
  \includegraphics[width=\columnwidth]{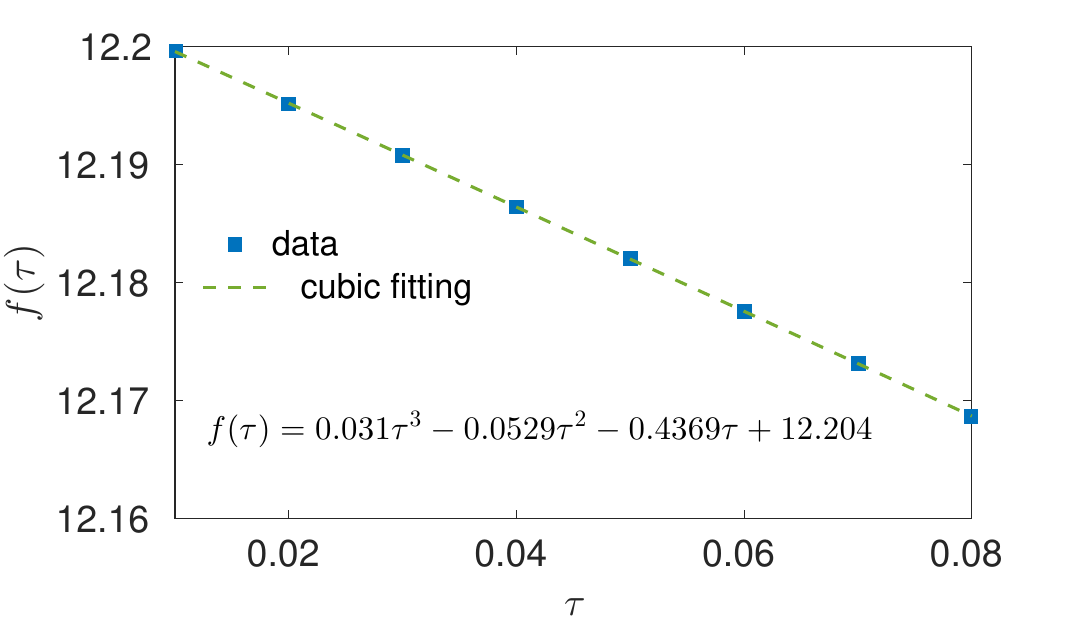}
  \caption{Polynomial fitting of the free energy $f(\tau)$.  $D^*=12$ and truncation multiplet dimension is $\chi^* = 650$  where the corresponding truncated bond dimension $\chi$ is larger than 1000. The linear coefficient is the variational energy $e_0=-0.4369$. All the fitting coefficients are accurate within 95\% confidence interval.}
  \label{fig:Efit}
\end{figure}

For the ground state energy $e_0=\langle\psi|H_\triangle+H_\triangledown)|\psi\rangle/N$, we expand the $\tau$-dependent free energy
\begin{equation}\label{eq:free}
\begin{split}
f(\tau)
=&\frac{1}{N}(\ln Z_\triangle(\tau)+\ln Z_\triangledown(\tau))	\\
=&f_0-\tau e_0+\mathcal{O}(\tau^2),
\end{split}
\end{equation}
where 
\begin{equation}
Z_{\triangle/\triangledown}(\tau)=\langle\psi|\exp(-\tau H_{\triangle/\triangledown})|\psi\rangle
\end{equation} 
is evaluated as the norm of uniform tensors. The polynomial fitting of the free energy $f(\tau)$ for TNS with multiplet dimension $D^*=12$ ($D=29$) is shown in Fig.~\ref{fig:Efit} and the ground state energy is $e_0=-0.4369$. Here, to compute the variational energy, we utilize the state-of-the-art SU(2) symmetry implementation\cite{symmetryLi,SymmetryLiu,SymmetrySingh,SymmetryWeichselbaum}. 

\begin{figure}[t]
  \centering
  \includegraphics[width=\columnwidth]{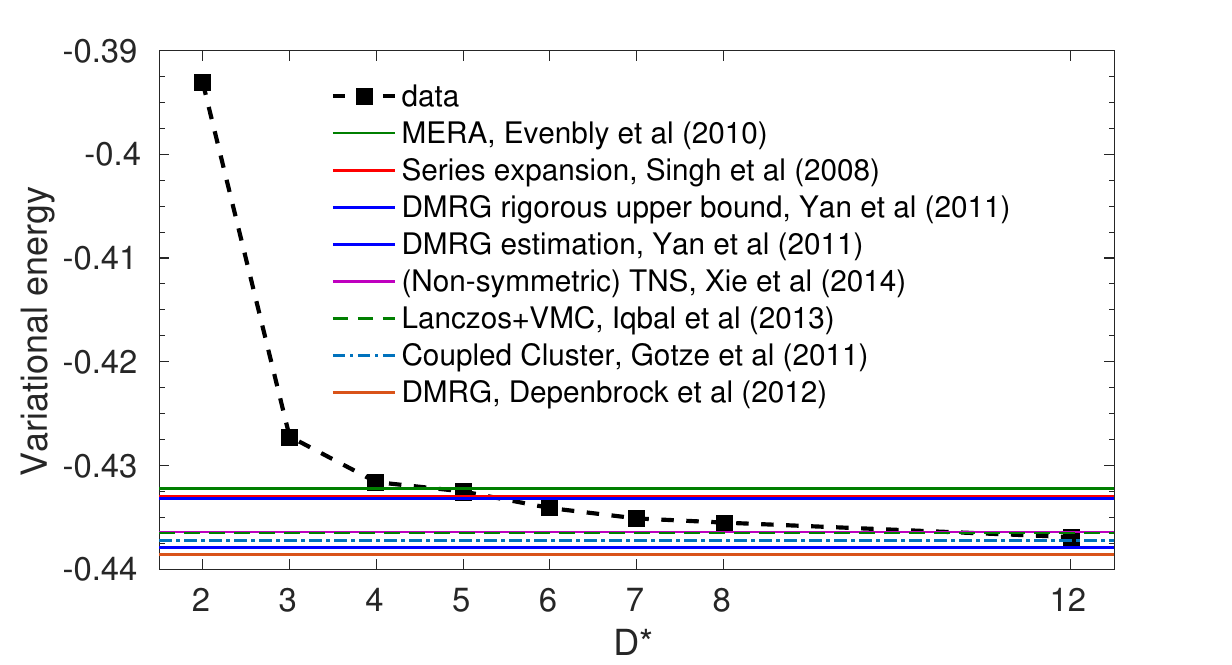}
  \caption{Variational energy of $S=1/2$ KHAFM for different multiplet dimensions $D^*$. The upper bonds on the ground-state energy obtained by multiscale entanglement renormalization ansatz (MERA)\cite{Evenbly2010}, the variational energies obtained by series-expansion methods\cite{Singh2008}, DMRG\cite{Yan2011,Depenbrock2012}, Lanczos improved variational Monte Carlo\cite{Iqbal2013}, coupled-cluster expansion\cite{Gotze2011} and non-symmetric TNS\cite{Xie2014}, are also shown for comparison.}
  \label{fig:ED}
\end{figure}

We provide our energy results mainly illustrated in Fig.~\ref{fig:ED}. For all the bond dimensions, we perform 10 steps TRG to obtain the energy. With increasing of the multiplet dimension $D^*$, the energy gradually decreases. The maximum multiplet dimension $D^*$ in our simulations is $D^*=12$ with variational energy per site $e_0 = -0.4369$. The energy comparison with other results is also displayed in Fig.~\ref{fig:ED}.

\section{Conclusion and Discussion}\label{eq.discussion}

In conclusion, we study the Kagome antiferrgomagnetic Heisenberg model through a symmetric tensor network approach. By keeping all the symmetries including $\mathrm{SU}(2)$ spin rotation, time reversal and translation symmetries, we obtain a symmetric TNS via a long imaginary time evolution. By computing modular matrices, we show that Kagome Heisenberg model is a gapped spin liquid with the $\mathbb{Z}_2$-topological order. From the system-size dependence of the modular matrices, we infer a long correlation length around 10 unit cell length implying a small finite gap. Furthermore, Our variational energy per site is $e_0 = -0.4369$ up to multiplet dimension $D^*=12$, which is 0.3\% higher than that of DMRG\cite{Depenbrock2012}. 

Experimentally, the spin gap in the Kagome spin liquid is confirmed in NMR\cite{Fu2015,Feng2017} and neutron scattering\cite{Han2016}. The fractional spin excitations (spinons) have been detected as fractionalized spin-wave continuum in neutron scattering measurements\cite{Han2012, Han2016}. Furthermore, the spin-1/2 quantum number of spinons is revealed in NMR measurements\cite{Feng2017}. The experimental evidence (e.g. gap and spin-1/2 quantum number) exclusively supports the $\mathbb{Z}_2$-gauge type (i.e. toric code type) topological order in the kagome spin liquid according to the theoretical argument\cite{ZaletelVishwanath2015}. With the help of SU(2) symmetry in our local tensors, we can use quantum transfer matrix\cite{Haegeman2015} to resolve the topological information (including non-universal information, such as correlation lengths) for low-energy excitations. %\HH{resolve the non-universal information for low-energy topological excitations, such as their energies.}
Fractionalized spin excitations in neutron experiment\cite{Han2012} have been interpreted in terms of bosonic spinons in Ref.~\onlinecite{Sachdev2014} and fermionic spinons in Ref.~\onlinecite{Mei2015a}, respectively. Bosonic and fermionic spinons have different symmetry fractionalizations and then are potentially resolved and verified in the neutron scattering experiments\cite{Mei2015a}.

\section{Acknowledgment}

We thank many helpful discussions with Nicolas Regnault, Roman Orus, Ling Wang, Zheng-Cheng Gu and Lukasz Cincio. 
JYC and HH thank the Perimeter Institute for Theoretical Physics for hospitality.
HH also thanks the support of the physics department of Princeton University. 
X.-G. Wen is supported by NSF Grant No. DMR-1506475 and NSFC 11274192. 
This research was supported in part by Perimeter Institute for Theoretical Physics. Research at Perimeter Institute is supported by the Government of Canada through the Department of Innovation, Science and Economic Development Canada and by the Province of Ontario through the Ministry
of Research, Innovation and Science.

\appendix
\clearpage

\section{Keeping Track of Symmetries}\label{app.symmetries}

There are two resources in TNS algorithms that violate the gauge symmetry and thus "kill" the topological order. The first one is the wrongly truncated singular vectors in truncated singular value decomposition (SVD). The second one is the floating-point round-off error in numerical SVD. Here we take a simple example to explain how to
get rid of these errors in the truncated SVD.

Suppose we have a two-leg tensor (matrix) $M$ with a symmetry $G$:
\begin{eqnarray}\label{eq:gauge}
\sum_{j_1,j_2} G_1(i_1, j_1)G_2(i_2,j_2)M(j_1,j_2)=M(i_1, i_2),
\end{eqnarray}
where $G_1,G_2\in G$ are on different legs. The truncated SVD is usually implemented to minimize the cost function $||M - \tilde{M}||_2$:
\begin{eqnarray}\label{eq:svd}
M = U S V^T\rightarrow\tilde{M}= \tilde{U} \tilde{S} \tilde{V}^T.
\end{eqnarray}
Here $\tilde{S}$ contains only the $D_{\text{cut}}$ largest singular values and $\tilde{U},\tilde{V}$ contain $D_{\text{cut}}$ corresponding singular vectors. Since $M$ has a symmetry $G$ as shown in Eq.~\eqref{eq:gauge}, the singular values are degenerate. The truncated bond dimension $D_{\text{cut}}$ should be carefully selected such that the truncated singular vectors contain all degenerate ones. 

If we keep all degenerate singular values up to a certain threshold, truncation error in SVD does not break any symmetry; i.e., the truncated singular vector space has the symmetry satisfying the following symmetry condition:
\begin{eqnarray}\label{eq:gaugeCond}
G_1 \tilde{U}\tilde{G} =\tilde{U},\quad \tilde{G}^{-1}\tilde{V}^TG_2=\tilde{V}^T.
\end{eqnarray}
Here $\tilde{G}$ is the symmetry matrix on the decomposed leg, and can be easily obtained from the symmetry condition Eq. \eqref{eq:gaugeCond}. Generally, $G_{1,2}$ are unitary and then we have the unitarity condition for symmetries on the decomposed bond, $\tilde{G}^\dag \tilde{G} = 1$. However, round-off error may violate the unitarity. After fixing the unitarity, we symmetrize the tensor $\tilde{U}$ and $\tilde{V}$ according to the symmetry condition Eq. (\ref{eq:gaugeCond}) and then round-off error is removed.

\section{Simple Update}\label{app.update}

In this appendix, we briefly introduce the simple update algorithm which is utilized in our work to find the ground states with TNS. For more details of simple update, we refer to Ref.~\onlinecite{Xie2014}. The basic idea of simple update is to use a relatively simple environment for the local tensors when truncating the bond dimension during imaginary-time evolution.
The algorithm goes as follows:

First of all, the imaginary-time evolution operator $\exp(-\tau H)$ is not a local operator although the Hamiltonian is made of local terms. It is an exponentially large matrix with respect to the system size. However, the evolution operator can be decomposed approximately according to the Trotter-Suzuki formula when $\tau \ll 1$:
\begin{eqnarray}
e^{-2\tau H}=e^{-\tau H_\triangle}e^{-\tau H_\triangledown}e^{-\tau H_\triangledown}e^{-\tau H_\triangle}+\mathcal{O}(\tau^3),
\end{eqnarray}
where $H=H_{\triangle}+H_{\triangledown}$; $H_\triangle$ and $H_\triangledown$ are the interactions defined, respectively, on all upward and downward triangles. And we apply the evolution operator $e^{-\tau H_\triangle}$, $e^{-\tau H_\triangledown}$, $e^{-\tau H_\triangledown}$ and $e^{-\tau H_\triangle}$ successively to TNS over many iterative steps. Since all the terms in $H_{\triangle}$ commute, $e^{-\tau H_\triangle}$ can be decomposed into a product of local evolution operators. Similarly for $e^{-\tau H_\triangledown}$. In this step, the symmetry between upward- and downward-oriented triangles are broken in the sense that the energies of upward- and downward-triangles may be different. This symmetry breaking can be fixed when we take a very small value of $\tau$ in every iteration.

A diagrammatic representation of the evolution $e^{-\tau H_\triangle}$ is shown in Fig. 1 (b) in the main text. After the local evolution operators act on local tensors, the products of tensors on the upward-triangles turn out to be $e^{-\tau H_\triangle} \mathrm{tTr}(T_1M^1M^2M^3)$ as shown in the left of Fig. 1 (b), which is a three-leg tensor with dimension $dD\times dD\times dD$. When we decompose this tensor to elementary ones and do truncation, the effect of the environment is included by introducing a positive bond vector $\lambda$ to mimic the environment (not shown in Fig. 1 (b)), which is square root of singular values from last iteration.
In Ref.~\onlinecite{Xie2014}, truncated high order singular value decomposition (T-HOSVD) is used to get new approximate tensors after evolution. However, since T-HOSVD is not the optimal truncation \cite{Kolda2009}, we use the high order orthogonal iteration (HOOI) method to do the truncation\cite{Kolda2009, Lathauwer2000}.

During the symmetric update, we decrease the Trotter time $\tau$ from $10^{-3}$ to $10^{-6}$. When $\tau$ is relative large, the singular value weights on up-triangle and down-triangle are slightly different. With decreasing $\tau$ , the difference diminishes less than $10^{-8}$. Also we find singular values weights have the $2\pi/3$ lattice rotation symmetry. Our states are symmetric within numerical error.

Although we use HOOI for the optimal truncation, the HOOI approximation is still not global approximation. More precisely, HOOI approximation is not the global approximation for the updated wave functions. It only involves local tensors. In simple update, the issue is considered by inserting diagonal matrices between nearest local tensors, which are used to simulate the environment of local tensors self-consistently\cite{Jiang2008a}.

Therefore, the problem with simple update is that it does not take precise account of the environment for local tensors which are updated in each iteration. Hence it may underestimate the long-range correlation or entanglement of the spins with small bond dimension. It is generally believed that more refined method like full update \cite{Jordan2008} can fix this issue.

The spin collection $J$ of $SU(2)$ spins on the virtual legs of local tensors may be influenced by the update method. Here we have utilized simple update to arrive at these bond dimensions and the corresponding gauge structure. We stress that they cannot be arbitrary spins in the following.

Because the local tensors are invariant under $SU(2)$ rotation, Eq.~(\ref{eq:symm}) in the main text, all the spins on the legs of local tensors should form a spin singlet. In other words, tensor $T_{1,2}$ and $M^{s_1,s_2,s_3}$ only contain the trivial representation of the tensor products of three $SU(2)$ representations whose coupling coefficients are known as $3j$ symbols. Hence, the set $J$ cannot be arbitrary collection of spins. The choice of $J$ must be able to form a spin singlet. 

For example, in the case of $D=3$, $J$ cannot be a spin-$1$, although the Hilbert space of spin-$1$ is $3$ dimensional. The reason is that: together with the physical spin-$1/2$, $M^{s_i}$ cannot form a spin singlet. See Eq.~(\ref{eq:symm}) and (\ref{eq:V}) in the main text. On the contrary, we need to select $\{0,\frac{1}{2}\}$ whose Hilbert space is also 3 dimensional.

\bibliography{kagomeZ2}
\end{document}